\documentclass[12pt]{article}
\usepackage{a4}
\usepackage{amsfonts,amssymb,amsmath,amsthm}
\usepackage{graphicx}
\usepackage{slashed}
\usepackage{cite}

\begin{document}
\title{Induced Chern-Simons action on noncommutative torus}
\author{D.~V.~Vassilevich\thanks{On leave from V.~A.~Fock Institute of
Physics, St.~Petersburg University, Russia.
E.mail:\ {\texttt{dmitry@dfn.if.usp.br}}}\\
{\it Instituto de F\'isica, Universidade de S\~ao Paulo,}\\ {\it
Caixa Postal 66318 CEP 05315-970, S\~ao Paulo, S.P., Brazil}}
\maketitle
\begin{abstract} We compute a Chern-Simons term induced by
the fermions on noncommutative torus interacting with two $U(1)$
gauge fields. For rational noncommutativity $\theta \propto P/Q$
we find a new mixed term in the action which involves only those
fields which are $(2\pi)/Q$ periodic, like the fields in a crystal
with $Q^2$ nodes.
 \end{abstract}
It is known since a long time that quantum one-loop corrections
due to 3-dimensional fermions generate the Chern-Simons action
\cite{Redlich,Alvarez-Gaume:1984nf,Niemi:1986kh}. This fact has
far reaching physical consequences, see \cite{Dunne:1998qy} for
a review. It is natural to consider this mechanism in the
framework of noncommutative (NC) field theorie. This was done in
\cite{Chu:2000bz,Grandi:2000av} on the 3-dimensional Moyal plane.
Different properties of NC Chern-Simons theories were studied
in a number of publications, see, e.g., \cite{Asano:2004hy} and
references therein.

The main novelty of the present work is that we consider the Dirac
fermions on a {\em compact} NC manifold (an NC 3-torus) which are
coupled to {\em two} independent $U(1)$ fields which act by left
and right Moyal multiplications. We calculate the parity-violating
part of the effective action (which generalizes the Chern-Simons
action for this case). We find a mixed contribution to this
action. For a rational NC parameter, $\theta/(2\pi)=P/Q$ this
mixed term exhibits a very interesting property: it involves
interactions only between the fields which are $(2\pi)/Q$-periodic
in the NC directions. We see a kind of dynamical crystallization
of the torus due to the effects of NC quantum field theory.

In this work we use the zeta-function regularization and the heat-kernel
methods (applied earlier to commutative low-dimensional fermionic
systems in e.g. \cite{GamboaSaravi:1984fz,GamboaSaravi:1994aq}).
Physical and mathematical aspects of this very powerful machinery
are reviewed in \cite{GKbook,Vreview}. For the Moyal-type noncommutativity
the heat kernel expansion was constructed in
\cite{Vassilevich:2003yz,Gayral:2004ww,Vassilevich:2005vk,Gayral:2006vd}.

The Moyal star-product on $\mathbb{T}^3$ is defined as usual by
\begin{equation}
f_1 \star f_2 (x) = \exp \left( \frac i2 \Theta^{\mu\nu}
\partial_\mu^x
\partial_\nu^y \right) f_1(x) f_2(y) \vert_{y=x}.\label{Mprod}
\end{equation}
The constant noncommutativity parameter $\Theta$ is an
antisymmetric $3\times 3$ matrix which is inevitably
degenerate\footnote{On noncompact NC manifolds degenerate $\Theta$
may cause problems in quantum theory \cite{Gayral:2004cu}. This is
one of the reasons to prefer the NC torus over the NC plane in
this work.}.
Our principle example is
\begin{equation}
\Theta^{12}=-\Theta^{21}\equiv \theta,\qquad \Theta^{13}=\Theta^{23}
=0.\label{mainex}
\end{equation}
Main formulae will be valid for generic $\Theta^{\mu\nu}$, but the
choice (\ref{mainex}) makes the results most transparent.

We take classical action for the Dirac fermions in the form
\begin{equation}
S=\int d^3x \sqrt{g} \bar\psi \slashed{D} \psi, \label{feract}
\end{equation}
where
\begin{equation}
\slashed{D}=i\gamma^\mu \left( \partial_\mu +iL(A_\mu^L)+ i
R(A_\mu^R) \right)\,. \label{Dirop}
\end{equation}
Here $L$ and $R$ are left and right Moyal multiplications,
$L(f)\phi =f\star \phi$, $R(f)\phi =\phi \star f$. Formal adjoints
of these operators coincide with multiplications by complex
conjugate functions, e.g.  $L(f)^\dag = L(f^*)$. It is convenient
to keep the ranges for all coordinates $x^\mu$ on the torus
$\mathbb{T}^3$ from $0$ to $2\pi$, but to allow for a constant
Euclidean metric $g_{\mu\nu}$. The $\gamma$-matrices are defined
by the condition
$\gamma^\mu\gamma^\nu+\gamma^\nu\gamma^\mu=2g^{\mu\nu}$, and ${\rm
tr} \gamma^\mu \gamma^\nu \gamma^\rho = 2i \epsilon^{\mu\nu\rho}$
with $\epsilon^{123}=g^{-1/2}$. To simplify our discussion we
impose periodic boundary conditions on the fermions on
$\mathbb{T}^3$. A short discussion of anti-periodic boundary
conditions is postponed until the end of this Letter.

As compared to
previous works on induced NC Chern-Simons theory
\cite{Chu:2000bz,Grandi:2000av} we have two independent vector
fields instead of one. The case $A_\mu^R=0$ corresponds to gauge
fields in the fundamental representation in the terminology of
\cite{Grandi:2000av} or to the Dirac fermions in the terminology of
\cite{Chu:2000bz}. $A_\mu^L=0$ corresponds to anti-fundamental
gauge fields \cite{Grandi:2000av}, and $A_\mu^R=-A_\mu^L$ is the
adjoint representation \cite{Grandi:2000av} or Majorana fermions
\cite{Chu:2000bz}. There are two gauge symmetries\footnote{General
discussion of gauge symmetris compartible with noncommutativity
can be found in \cite{Chaichian:2001mu}.}
corresponding to
two $U(1)$ gauge fields (so that we have a double gauging
of $U(1)$ \cite{Liao:2004mu}):
\begin{eqnarray}
&&\psi \to U_L \star \psi \star U_R,\qquad \bar \psi \to
U_R^{\dag}
\star \bar \psi \star U_L^{\dag},\label{gtran}\\
&&iA_\mu^L\to U_L \star \partial_\mu U_L^{-1} +i U_L \star A_\mu^L
\star U_L^{-1} ,\nonumber\\
&&iA_\mu^R \to\partial_\mu U_R^{-1}\star U_R +i U_R^{-1} \star
A_\mu^R \star U_R,\nonumber
\end{eqnarray}
where $U_{L,R}$ are star-unitary, $U_{L,R}\star U_{L,R}^\dag =1$.
The Dirac operator is transformed as $\slashed{D}\to R(U_R)
L(U_L)\slashed{D} L(U_L^{-1}) R(U_R^{-1})$. Because of these two symmetries
there are two independent currents,
$j^\mu_L=\psi^b \star \bar\psi^a \gamma^\mu_{ab}$ and
$j^\mu_R=\bar\psi^a \star \psi^b \gamma^\mu_{ab}$ (with $a,b$ being
spinor indices) which are separately covariantly conserved
\begin{equation}
\partial_\mu j_L^\mu + iA_\mu^L \star j_L^\mu - ij_L^\mu \star A_\mu^L=0,
\quad
\partial_\mu j_R^\mu - iA_\mu^R \star j_R^\mu + ij_R^\mu \star A_\mu^R=0.
\label{cons}
\end{equation}
The existence of two independent vector currents is an additional
motivation to introduce two vector field coupled to them, which is
even necessary if one uses such vectors to study the dynamics of
collective excitations of the fermions.

The Dirac operator squared is an operator of Laplace type, i.e.
\begin{eqnarray}
&&\slashed{D}^2=-(\nabla^2 + E),\nonumber\\
&&\nabla_\mu =\partial_\mu +iL(A_\mu^L)+
i R(A_\mu^R) ,\label{DirLap}\\
&&E=\frac i2 [\gamma^\mu,\gamma^\nu] \left( L(\partial_\mu A_\nu^L +
i A_\mu^L\star A_\nu^L) + R(\partial_\mu A_\nu^R +i A_\nu^R \star A_\mu^R)
\right).\nonumber
\end{eqnarray}

We are going to use the heat-kernel methods, so let us remind some
basic facts \cite{Gayral:2006vd} regarding the heat kernel
expansion on NC torus for an operator of Laplace type containing
both left and right Moyal multiplications. The net result of
\cite{Gayral:2006vd} is that the heat kernel expansion looks
precisely as in the commutative case if one uses a modified trace
operation. To define this trace we need a little bit of number
theory. The construction will be given in all detail for the
special case of (\ref{mainex}). For generic $\Theta^{\mu\nu}$ the 
reader can consult \cite{Gayral:2006vd}.
A real number $\alpha$ is called Diophantine if there are
two positive constants $C$ and $\beta$ such that
\begin{equation}
\inf_{P\in \mathbb{Z}} | \alpha Q - P |\ge \frac C{|Q|^{1+\beta}}
\qquad \mbox{for all}\quad Q\in \mathbb{Z}\,.\label{Dioph}
\end{equation}
In other words, the Diophantine numbers are the numbers which
cannot be too well approximated by rational numbers. We suppose
that $(\theta/2\pi)$ is either rational, or
Diophantine\footnote{If $(\theta/2\pi)$ is neither this nor that,
the heat kernel asymptotics are unstable, i.e. the powers of the
proper time appearing in the asymptotic expansion depend on the
ultra-violet behavior of the background fields, see App.\ B of
\cite{Gayral:2006vd}. The fact that quantum field theory on NC
torus is very sensitive to the number theory nature of $\theta$
was noted already in \cite{Krajewski:1999ja}.}.

Next we define a special subset $\mathcal{Z}$ of the Fourier momenta.
A momentum $q\in \mathbb{Z}^3$ belongs to $\mathcal{Z}$ iff
\begin{equation}
(2\pi)^{-1} \Theta q \in \mathbb{Z}^3.
\label{defZ}
\end{equation}
(This formula is also valid for generic $\Theta^{\mu\nu}$).
Let us give some examples. In the commutative case, $\theta=0$,
the condition (\ref{defZ}) is satisfied by all momenta, and
$\mathcal{Z}=\mathbb{Z}^3$. If $\theta/(2\pi)$ is irrational (then
under our assumptions it is also Diophantine), only the $q_3$ can
be non-zero, and $\mathcal{Z}=\{ 0 \}\otimes \{ 0 \} \otimes
\mathbb{Z}$. The most interesting case is rational
noncommutativity. Then
\begin{equation}
\mathcal{Z}=Q\cdot \mathbb{Z} \otimes Q\cdot \mathbb{Z} \otimes \mathbb{Z}
\qquad \mbox{for} \qquad \theta/(2\pi)=P/Q \label{ratthet}
\end{equation}
with $P\in \mathbb{Z}$ and $Q\in \mathbb{N}$. (Of course, $P/Q$ must be
irreducible). The set $\mathcal{Z}$
depends on $Q$ but not on $P$. Note that two previous
cases may be obtained by taking formal limits $Q\to 1$ and $Q\to\infty$
in (\ref{ratthet}), respectively. In the sense of this remark $\mathcal{Z}$
is uniquely defined by a number $Q\in \mathbb{N}\cup \{\infty\}$.

Let us now define the trace. Consider an operator which can be represented
as a product of left and right Moyal multiplications
$L(l)R(r)$, possibly with some matrix structures. Then
\begin{equation}
{\rm Sp}(L(l) R(r))=\sqrt{g} \sum_{q\in \mathcal{Z} }
\tilde l(-q) \, \tilde r(q) \,,\label{defSp}
\end{equation}
where $\tilde l$ and $\tilde r$ are the Fourier modes:
\begin{equation}
\tilde l(k)=(2\pi)^{-3/2} \int d^3x \, l(x)e^{-ikx}.\label{Four}
\end{equation}
The definition (\ref{defSp}) includes also trace over all matrix
indices, which we do not write explicitly here.

Now consider an operator $P$ on NC $\mathbb{T}^n$ which can be represented as
$P=-(\nabla^2+E)$ where $E$ and $\omega_\mu$ in $\nabla_\mu =
\partial_\mu +\omega_\mu$ are zeroth order operators, i.e.,
$E$ and $\omega_\mu$ are combinations of left and right Moyal
multiplications. Such operators are called generalized Laplacians
($\slashed{D}^2$ is an example, see eq.\ (\ref{DirLap})). It was
demonstrated in \cite{Gayral:2006vd} that for such operators the
heat operator $e^{-tP}$ exists for positive $t$ and is trace
class, and there is a full asymptotic series as $t\to +0$
\begin{equation}
{\rm Tr}\left( L(l)R(r)e^{-tP}\right) \simeq \sum_{m=0}^{\infty}
t^{(n-m)/2} a_{2m} (L(l)R(r),P)\,.\label{asymptotex}
\end{equation}
${\rm Tr}$ is the $L_2$ trace. In particular, first couple of the
heat kernel coefficients\footnote{We use ''inflated notations''
for the heat kernel coefficients, so that only even-numbered coefficients
appear usually on manifolds without a boundary. In this nomenclature
no half-integer indices appear also in the presence of boundaries,
see \cite{GKbook,Vreview}.} read
\begin{eqnarray}
&&a_0=(4\pi)^{-n/2} {\rm Sp} (L(l)R(r)),\label{a0}\\
&&a_2=(4\pi)^{-n/2} {\rm Sp} (L(l)R(r)E).\label{a2}
\end{eqnarray}
It is easy to figure out how the eqs.\ (\ref{defZ}) and (\ref{Four}) must
be generalized to arbitrary $n$.

In this Letter we employ the zeta-function regularization which a
is a proper instrument to keep gauge invariance throughout the
calculations \cite{Deser:1997nv}. We first use the relation
\cite{Alvarez-Gaume:1984nf,Niemi:1986kh,Deser:1997nv} between the
parity-violating part of the effective action and the eta
invariant
\begin{equation}
\Gamma^{\rm pv}=i\frac {\pi}2 \eta (0),
\label{effeta}
\end{equation}
which is defned through a sum over the eigenvalues of $\slashed{D}$,
\begin{equation}
\eta (s)=\sum_{\lambda_n>0} (\lambda_n)^{-s} -\sum_{\lambda_n<0}
(-\lambda_n)^{-s}.\label{defzeta}
\end{equation}
This spectral function measures the spectral asymmetry of the Dirac operator.
Next couple of steps repeat quite literally those of
\cite{Alvarez-Gaume:1984nf}.
We make use of an integral representation of the
eta function and replace the sum over the spectrum by ${\rm Tr}$.
\begin{equation}
\eta (s)=\frac 2{\Gamma \left( (s+1)/2\right)} \int_0^\infty d\tau\,
\tau^s \, {\rm Tr}\left( \slashed{D} e^{-\tau^2\slashed{D}^2} \right) .
\label{inteta}
\end{equation}
Let us vary $A_\mu^{L,R}$ in $\slashed{D}$. The variation of $\eta(s)$
reads
\begin{equation}
\delta \eta (s)=\frac 2{\Gamma \left( (s+1)/2\right)} \int_0^\infty d\tau\,
\tau^s \, \frac d{d\tau} {\rm Tr}
\left((\delta \slashed{D}) \tau e^{-\tau^2\slashed{D}^2} \right) .
\label{vareta}
\end{equation}
Now, by taking $s\to 0$ (and assuming that the heat kernel decays fast enough
at $\tau^2\to \infty$, which is usually true) one arrives at
\begin{eqnarray}
\delta \eta (0)&=&-\frac 2{\sqrt{\pi}} \lim_{\tau\to 0}
{\rm Tr} \left( (\delta \slashed{D}) \tau
e^{-\tau^2\slashed{D}^2} \right)\nonumber\\
&=&-\frac 2{\sqrt{\pi}} \lim_{t\to 0}
{\rm Tr} \left( (\delta \slashed{D}) t^{1/2}
e^{-t\slashed{D}^2} \right).\label{detahk}
\end{eqnarray}
To evaluate this limit we use the heat kernel expansion (\ref{asymptotex}).
The coefficient $a_0$ does not contribute because of the $\gamma$-trace.
We are left with
\begin{equation}
\delta \eta (0)=-\frac 2{\sqrt{\pi}} a_2(\delta \slashed{D},\slashed{D}^2).
\label{detaa2}
\end{equation}
This heat kernel coefficient reads
\begin{equation}
a_2(\delta \slashed{D},\slashed{D}^2)=\frac 1{8\pi^{3/2}} {\rm
Sp}_2 \left( \gamma^\mu (L(-\delta A_\mu^L)+R(-\delta
A_\mu^R))\cdot E \right), \label{a22}
\end{equation}
where $E$ is given by (\ref{DirLap}) and the subscript "$2$" in ${\rm Sp}_2$
reminds to calculate the trace over the spinor indices which yields
\begin{eqnarray}
&&a_2(\delta \slashed{D},\slashed{D}^2)=\frac 1{4\pi^{3/2}}\,
\epsilon^{\mu\nu\rho} {\rm Sp} \left( (L(\delta A_\mu^L)+R(\delta
A_\mu^R))\right.\nonumber\\
&&\qquad \cdot\left. (L(\partial_\nu A_\rho^L +i A_\nu^L \star
A_\rho^L)+R(\partial_\nu A_\rho^R +i A_\rho^R \star A_\nu^R)
\right).\label{a23}
\end{eqnarray}

To calculate the remaining trace we need a couple of relations.
 First,
\begin{equation}
{\rm Sp} (L(f))={\rm Sp} (R(f))=\int d^3 x\,\sqrt{g}\, f(x)
\label{SpL}
\end{equation}
where to apply the definition (\ref{defSp}) one has to write
$L(f)=L(f)R(1)$. Next, we have the symmetry property
\begin{eqnarray}
&&{\rm Sp} (L(f) R(f_1\star f_2))={\rm Sp} (L(f) R(f_2 \star
f_1))\nonumber\\
&&{\rm Sp} (L(f_1\star f_2)R(f))={\rm Sp} (L(f_2 \star f_1)R(f))
\label{tracepro}
\end{eqnarray}
for any functions $f$, $f_1$, $f_2$. The relations (\ref{SpL}) and
(\ref{tracepro}) were derived in \cite{Gayral:2006vd}.

To derive another useful property we should first define a
$Q$-periodic projection of functions on $\mathbb{T}^3$
\begin{equation}
[f]_Q:=(2\pi)^{-3/2} \sum_{k\in\mathcal{Z}} \tilde f(k) e^{ikx}.
\label{perpro}
\end{equation}
(For generic $\Theta^{\mu\nu}$ one can define a projection
$[f]_{\mathcal{Z}}$; the right hand side of (\ref{perpro}) remains unchanged).
This is indeed a projection, $[[f]_Q]_Q=[f]_Q$. For a rational
$\theta/(2\pi)=P/Q$ this operation selects
a part of $f$ which is periodic in $x^1$ and $x^2$ coordinates with the
period $2\pi/Q$. In the irrational (Diophantine) case this operation
(which may be denoted as $[f]_\infty$ according to the remark below eq.\
(\ref{ratthet})) selects just the average value of $f$ on the
$\mathbb{T}^2$ spanned by $x^1$ and $x^2$. In the commutative case,
$\mathcal{Z}=\mathbb{Z}^3$, this is the identity map $[f]_1 =f$.
By making the Fourier transform back and forth one can prove
that
\begin{equation}
{\rm Sp} (L(l)R(r))=\int_{\mathbb{T}^3} d^3x \, \sqrt{g}\,
[l]_Q \cdot [r]_Q=\int_{\mathbb{T}^3} d^3x \, \sqrt{g}\,
[l]_Q \star [r]_Q\,.\label{Spro}
\end{equation}

By using (\ref{SpL}), (\ref{tracepro}) and (\ref{Spro}) we rewrite
(\ref{a23}) as
\begin{eqnarray}
&&a_2(\delta \slashed{D},\slashed{D}^2)=a_2^L+a_2^R+a_2^{\rm
mixed},\label{a2fin}\\
&&a_2^L=4\pi^{-3/2} \int d^3 x\,\sqrt{g}\, \epsilon^{\mu\nu\rho}
(\delta A_\mu^L) \left( \partial_\nu A_\rho^L
+iA_\nu^L\star A_\rho^L\right)\,,\nonumber\\
&&a_2^R=4\pi^{-3/2} \int d^3 x\,\sqrt{g}\, \epsilon^{\mu\nu\rho}
(\delta A_\mu^R) \left( \partial_\nu A_\rho^R
-iA_\nu^R\star A_\rho^R\right)\,,\nonumber\\
&&a_2^{\rm mixed}=4\pi^{-3/2} \int d^3 x\,\sqrt{g}\,
\epsilon^{\mu\nu\rho} \left( [(\delta A_\mu^L)]_Q  \partial_\nu
[A_\rho^R]_Q +[(\delta A_\mu^R)]_Q \partial_\nu [A_\rho^L]_Q
\right)\,.\nonumber
\end{eqnarray}
Cubic terms in $a_2^{\rm mixed}$ vanish due to (\ref{tracepro}).

In the zeta function regularization (see, e.g.
\cite{GamboaSaravi:1984fz}) the parity-violating part of the
effective action $\Gamma^{\rm pv}$ is expressed through the
Chern-Simons action $S_{\rm CS}$ by means of the relation
$\Gamma^{\rm pv}=\frac 12 S_{\rm CS}$. (In the Pauli-Villars
regularization, for examples, the PV masses also enter this
relation). We combine (\ref{effeta}) and (\ref{detaa2}) with
(\ref{a2fin}) to obtain a generalized Chern-Simons action induced
by fermionic fluctuations on the NC torus
\begin{eqnarray}
&&S_{\rm CS}=S_{\rm CS}^L+S_{\rm CS}^R + S_{\rm CS}^{\rm mixed},
\label{SCS}\\
&&S_{\rm CS}^L=-\frac i{4\pi} \int d^3x \, \sqrt{g}
\epsilon^{\mu\nu\rho} \left( A_\mu^L\partial_\nu A_\rho^L
+\frac{2i}3 A_\mu^L \star A_\nu^L \star A_\rho^L
\right),\label{Sl}\\
&&S_{\rm CS}^R=-\frac i{4\pi} \int d^3x \, \sqrt{g}
\epsilon^{\mu\nu\rho} \left( A_\mu^R\partial_\nu A_\rho^R
-\frac{2i}3 A_\mu^R \star A_\nu^R \star A_\rho^R
\right),\label{Sr}\\
&&S_{\rm CS}^{\rm mixed}=-\frac i{2\pi} \int d^3x \, \sqrt{g}
\epsilon^{\mu\nu\rho}  [A_\mu^L]_Q\partial_\nu
[A_\rho^R]_Q\,.\label{Smixed}
\end{eqnarray}
Note, that the combination $\sqrt{g}\epsilon^{\mu\nu\rho}$ does
not depend on the metric, so that $S_{\rm SC}$ is also
metric-independent (topological) as in the commutative case.

The "planar" terms (\ref{Sl}) and (\ref{Sr}) are pretty much
standard NC generalizations of the Chern-Simons action in all
particular cases considered in \cite{Chu:2000bz,Grandi:2000av}
(cf. also the discussion above eq.\ (\ref{gtran})). 
The reason is that the analytic expressions
for planar heat kernel coefficients are not sensitive to the
space-time topology. There is one particular case, $A^L=A^R$,
when all noncommuativity effects disappear from the planar
contributions. However, there is no $U(1)$ subgroup of
the gauge transformations (\ref{gtran}) which respects
the choice $A^L=A^R$, so this case is not interesting.

The mixed term (\ref{Smixed})
exhibits a rather interesting physical property. It involves
interactions only between those background fields $A_\mu^{L,R}$ which
are $(2\pi)/Q$ periodic in the $x^1$ and $x^2$ directions. Such
fields remind us of solid state physics and correspond to a
crystal consisting of $Q\times Q$ fundamental domains. By
stretching a bit the terminology we can say that the two-torus is
dynamically crystallized due to NC quantum effects.
It would be very interesting to establish precise relations
between this effect and the works on the equivalence 
of field theories on NC torus with ratioanal noncommutativity
to matrix models (see, e.g., \cite{matrix}).

When using the spectral geometry methods one should not worry too
much about the gauge invariance since the eta function
(\ref{defzeta}) is manifestly gauge invariant. One can also check
the gauge invariance of (\ref{Sl}) - (\ref{Smixed}) by direct
calculations. To this end the relation (\ref{tracepro}) is very
useful.

The Chern-Simons action (\ref{SCS}) does not depend smoothly on
$\theta$. Moreover, we have calculated this action for a rational
or Diophantine noncommutativity only. Nevertheless, there is a
well defined commutative limit, though not an obvious one. Instead
of taking $\theta\to 0$, we take a rational NC parameter with
$Q=1$. Then the star-product becomes commutative (one easily gets:
$e^{ikx}\star e^{iqx}=e^{iqx}\star e^{ikx}=\pm e^{i(k+q)x}$),
though still not the ordinary one (unless $P$ is even). Let us
introduce two new vector fields by $A^L=\frac 12 (B+C)$,
$A^R=\frac 12 (B-C)$. We have
\begin{equation}
S_{\rm CS}\vert_{Q=1}=-\frac i{4\pi} \int d^3x \, \sqrt{g}
\epsilon^{\mu\nu\rho} B_\mu \partial_\nu B_\rho .\label{Scom}
\end{equation}
We see, that, precisely as one would expect in the commutative
limit, the field $C_\mu$ disappears, and the action for $B_\mu$ is
the standard abelian Chern-Simons action.

To be able to use the heat kernel asymptotics from
\cite{Gayral:2006vd} we assumed that $\psi$ is periodic on the
torus. If instead one chooses (perhaps physically more preferable)
anti-periodic boundary conditions in some of the coordinates, then
allowed fermionic momenta in corresponding directions are shifted
by $1/2$ to form a set which we call $\widetilde {\mathbb{Z}}^3$
instead of $\mathbb{Z}^3$ considered above. The definition
(\ref{defZ}) should be modified: $q\in \mathcal{Z}$ iff $q_\mu
\theta^{\mu\nu} k_\nu$ is $(2\pi)$ times an integer number for all
$k\in \widetilde{\mathbb{Z}}^3$. (Note that $\mathcal{Z} \subset
\mathbb{Z}^3$ as before since $q$'s are the Fourier momenta of 
bosonic background fields). This is the most important
modification, which, of course, also leads to some changes in the
periodic projections. Qualitatively our results remain unchanged.
We hope to consider anti-periodic boundary conditions in more
detail in a future publication.

To summarize, in this Letter we calculated the Chern-Simons action
induced by the parity anomaly of fermions on NC 3-torus. Due to
the compactness\footnote{To see that the compactness is essential
it is enough to compare the heat kernel expansions on the Moyal
plane \cite{Vassilevich:2005vk} and on the Moyal torus
\cite{Gayral:2006vd}.} of the manifold and due to the presence of
two independent gauge fields we found that there is a mixed
(non-planar) contribution to the action of a rather particular
form (\ref{Smixed}). This mixed term depends only on the fields
which are $(2\pi)/Q$-periodic and thus remind us of the fields in
a crystal. It would be interesting to test this result at physical
applications of the NC Chern-Simons, e.g. at the model
\cite{Susskind} of Quantum Hall Fluids.

\end{document}